\documentclass[%
reprint, prl,
superscriptaddress,
 amsmath,amssymb,
 aps,
]{revtex4-2}

\usepackage[T1]{fontenc}
\usepackage[
bookmarks=true,
colorlinks,
linkcolor=blue,
urlcolor=blue,
citecolor=olive,
plainpages=false,
pdfpagelabels,
final,
breaklinks=true
]{hyperref}

\usepackage{graphicx}
\usepackage{dcolumn}
\usepackage{bm}
\usepackage[dvipsnames]{xcolor}
\usepackage{physics}
\usepackage{pifont}
\usepackage{bbm}
\usepackage{multirow}
\usepackage{microtype}
\usepackage[normalem]{ulem}

\allowdisplaybreaks

\begin{document}

\preprint{APS/123-QED}

\title{Attosecond Quantum Optical Spectroscopy of Topological Materials}

\author{C.~Granados}%
\email{cagrabu@eitech.edu.cn}
\affiliation{Eastern Institute of Technology, Ningbo 315200, China}

\author{J.~Rivera-Dean}
\email{javier.dean@ucl.ac.uk}
\affiliation{Department of Physics and Astronomy, University College London, Gower Street, London WC1E 6BT, United Kingdom}

\author{D.~Kim}
\affiliation{Department of Materials Science and Engineering, Pohang University of Science and Technology, Pohang 37673, Korea}

\author{A.~Chac\'on}
\affiliation{Departamento de F\'isica, \'Area de F\'isica, Universidad de Panam\'a,
Ciudad Universitaria 3366 Octavio Mendez Pereira, Panama}

\author{W. Gao}%
\affiliation{Eastern Institute of Technology, Ningbo 315200, China}

\author{P.~Stammer}%
\email{philipp.stammer@icfo.eu}
\affiliation{ICFO—Institut de Ciencies Fotoniques, The Barcelona Institute of Science and Technology, 08860 Castelldefels (Barcelona), Spain}%
\affiliation{Atominstitut, Technische Universität Wien, 1020 Vienna, Austria}

\date{\today}
\begin{abstract}
High-harmonic generation (HHG) provides an ultrafast probe of quantum materials, yet
conventional spectral observables do not generally furnish a common signature of band topology. Here we show that quantum-light-driven HHG opens an additional information channel through the photon statistics of the emitted harmonics. Using the Haldane Chern insulator and the time-reversal-invariant Kane–Mele topological insulator, we calculate helicity-resolved equal-time intensity correlations and identify a common enhancement of high-order harmonic correlations in the topological phases. For the Haldane model, a dichroic correlation constructed from opposite helicity channels reverses sign with the Chern number. These signatures originate from the distinct nonlinear response sampled by quantum fluctuations of the driving field and are absent from the corresponding mean harmonic spectra as a common classifier. Our results establish photon-correlation spectroscopy as a route to interrogate topological matter with quantum light and provide a bridge between strong-field solid-state physics and quantum optics.
\end{abstract}

\maketitle

\emph{Introduction}---Light-matter interactions fundamentally shape our understanding of nature at short temporal and small spatial scales~\cite{Ivanov1}. They allow us to develop techniques in which light is used to control matter~\cite{Valeytronics} and to demonstrate that matter, in turn, affects light after a strong interaction~\cite{CatStates}. Moreover, the study of condensed matter systems is central to the understanding of fundamental physical phenomena~\cite{Vampa1, LightInduced, Ginzburg1}, and their application to modern technologies~\cite{BlueLed}. Conventional solid-state materials are classified as what is nowadays known as a topologically-trivial state~\cite{Chen1}: their band structures can be continuously deformed into flat bands without closing the direct energy gap, and their associated topological invariants vanish. As a consequence, their electronic properties are largely governed by the local band dispersion and can be readily modified by external fields~\cite{YuCardona2010}.

In contrast, topological matter~\cite{Thouless1,Haldane2} exhibits intrinsically different properties that are characterized by topological invariants~\cite{Chen1,Haldane2}. Experimentally, these exotic phases were first revealed by the discovery of the integer~\cite{Inte} and fractional~\cite{Fract} quantum Hall effects. Prototypical examples of topological systems include topological insulators (TIs), Chern insulators (CIs), Weyl semimetals~\cite{Bansil2016}, and quantum spin liquids~\cite{Savary2017}. In particular, CIs are characterized by a nonzero topological invariant, the Chern number, and by broken time-reversal symmetry~\cite{Haldane, QAHE}. The band structures of these systems exhibit a nontrivial Berry-curvature distribution across the Brillouin zone (BZ), leading to a quantized Hall conductivity in the absence of an external magnetic field~\cite{Haldane}. The non-vanishing Chern number ($C = \pm 1$ in the simplest case) reflects a globally twisted structure of the Bloch wavefunctions and guarantees the existence of robust chiral edge states~\cite{HasanKane2010, Thouless1}. Unlike trivial insulators, CIs cannot be adiabatically connected to a flat-band limit without closing the bulk gap, emphasizing their fundamental topological origin~\cite{Haldane}. The quintessential realization of a CI is the Haldane model (HM)~\cite{Haldane, Haldane2}, which describes spinless fermions on a honeycomb lattice with complex next-nearest-neighbor hopping amplitudes. Its time-reversal-invariant counterpart is the Kane-Mele model (KMM)~\cite{KMM}, which incorporates the electron spin through an intrinsic spin-orbit interaction. The KMM can be understood as two copies of the Haldane model with opposite Chern numbers for opposite spin sectors, resulting in a vanishing net Chern number but a nontrivial $\mathbb{Z}_2$ topological invariant~\cite{KMM,DasolKMM}. Consequently, it supports topologically protected edge states while preserving time-reversal symmetry, making it the prototypical model of a two-dimensional TI.

The convergence of ultrafast phenomena and topological matter was realized through high-harmonic generation (HHG)~\cite{Zong2023}, a highly nonlinear, non-perturbative process~\cite{HHG_theory}. It has been shown theoretically that HHG can discriminate between topologically trivial and nontrivial phases when driven by circularly polarized light (CPL)~\cite{Alexis1, Silva1}. Owing to its sensitivity to interband coherence, the anomalous velocities and the corresponding Berry curvature are encoded in the HHG spectrum. This allows one to learn about the underlying band topology and exhibits distinct signatures: right- and left-CPL drivers produce different helicities~\cite{Silva1}, and a circular dichroism that can be mapped onto the topological invariant~\cite{Alexis1}. This distinction was demonstrated experimentally in the prototypical three-dimensional TI bismuth selenide~\cite{PhaseTrans}. The reliability of HHG as a probe of topological phases nevertheless remains under active debate, particularly regarding its ability to unambiguously capture topological invariants and to cleanly resolve the spectral features associated with different phases~\cite{Lorenzo_HHG_Topo, Ofer1}.

However, all these studies have considered classical coherent laser sources. Recently, there is the emergence of quantum optical analysis of HHG~\cite{stammer2025colloquium, stammer2023quantum, lewenstein2021generation, cruz2024quantum}, allowing us to understand the consequences that non-classical driving fields, such as BSV~\cite{gorlach2023high, Javi1, petrovic2026generation, rivera2026emergence}, has on the ultrafast electron dynamics~\cite{Javi1,PS2023}. Furthermore THE BSV light was recently used to drive HHG in a solid state system~\cite{rasputnyi_HHG}. This opens the route to probe solid-state materials beyond classical fields, yet it has not been studied in the interaction with topological materials. Owing to the broader photon number distribution of BSV light compared to classical coherent laser of the same mean intensity, most pulses carry far fewer photons than the mean, while rare high photon number events can still drive HHG without accumulating damage, granting access to electron dynamics at pulse intensities far beyond the damage-threshold limit~\cite{rasputnyi_HHG}. Moreover, because of the strong subcycle quantum fluctuations of the field~\cite{stammer2026attosecond}, the BSV-driven electron and hole dynamics unfold in a coherent superposition of quasi-momentum states, in contrast to the well-defined quasi-momentum states of the quasi-particles generated by a coherent driver~\cite{Alexis1}. The resulting harmonic photon-number distribution hence departs from the Poissonian distribution characteristic of a coherent state~\cite{rivera2026attosecond}. These findings highlight the unique role of quantum light as a tool to probe states of matter under new conditions. To date, however, all theoretical studies of quantum-light-driven HHG have been restricted to trivial materials~\cite{gorlach2023high,Javi1,petrovic2026generation,rivera2026emergence,QL_Solids,stammer2026fluctuation}.

In this Letter, we investigate the nonlinear response of topological matter to bright quantum light, a regime where quantum fluctuations and photon statistics become relevant. In particular, we study two different topological systems (i) a CI, represented by the HM, and (ii) a TI, represented by the KMM. By computing the equal-time intensity autocorrelation function $g^{(2)}(0)$, we show that quantum light allows distinguishing the topological from the trivial matter. Most importantly, we show that the two distinct classes of topological material considered here respond in the same way to quantum light. Our findings open a new route towards a more general characterization of topological matter beyond classical driving fields, connecting quantum optics and topological matter with ultrafast laser physics.

\emph{Theoretical framework}---The quantum light effects on the HM and KMM are described within the generalized positive $P$ representation of displaced squeezed vacuum states, following the approach of Refs.~\cite{Javi1,QL_Solids}. In short, the total HHG spectrum resulting from the different squeezed states is an average of the individual spectra over a marginal of the Husimi function~\cite{Javi1}, with the HHG process driven by circularly polarized quantum light in the initial state
\begin{equation}\label{eq:state_initial}
        \ket{\Phi_{\omega}(t_0)} =  \hat{D}_{\omega,\perp}(\alpha_{\omega}) \hat{S}_{\omega,\perp}(r) \ket{0_\perp}
             \otimes \hat{D}_{\omega,\parallel}(i \alpha_{\omega}) \ket{0_\parallel},    
\end{equation}
where $\hat{D}_{\omega,\mu}(\alpha)$ and $\hat{S}_{\omega,\mu}(r)$ are the displacement and squeezing operators acting on the field mode with polarization component $\mu = \{ \perp, \parallel\}$, respectively. Here we consider the squeezing to be applied to one of the linearly polarized components of the $\omega$ driver. In particular, we numerically evaluate the expectation value of the charge current in the solid, which is obtained as
\begin{equation}
    J(t) = \int Q(\alpha) J[A(t;\alpha)]d^2\alpha, 
\end{equation}
where $J[A(t;\alpha)]$ is calculated from the time-dependent density matrix (TDDM) formalism, using the classical vector potential associated with the coherent state $|\alpha\rangle$, i.e., $A(t;\alpha) = \bra{\alpha}\hat{A}(t)\ket{\alpha}$, with $\hat{A}(t)$ the vector potential operator of the driving field. The interband and intraband currents are also evaluated independently for each field value and subsequently averaged over the Husimi distribution.

\begin{figure}[h!]
\centering 
\includegraphics[width=1\linewidth,trim=0cm 0cm 0cm 0cm,clip]{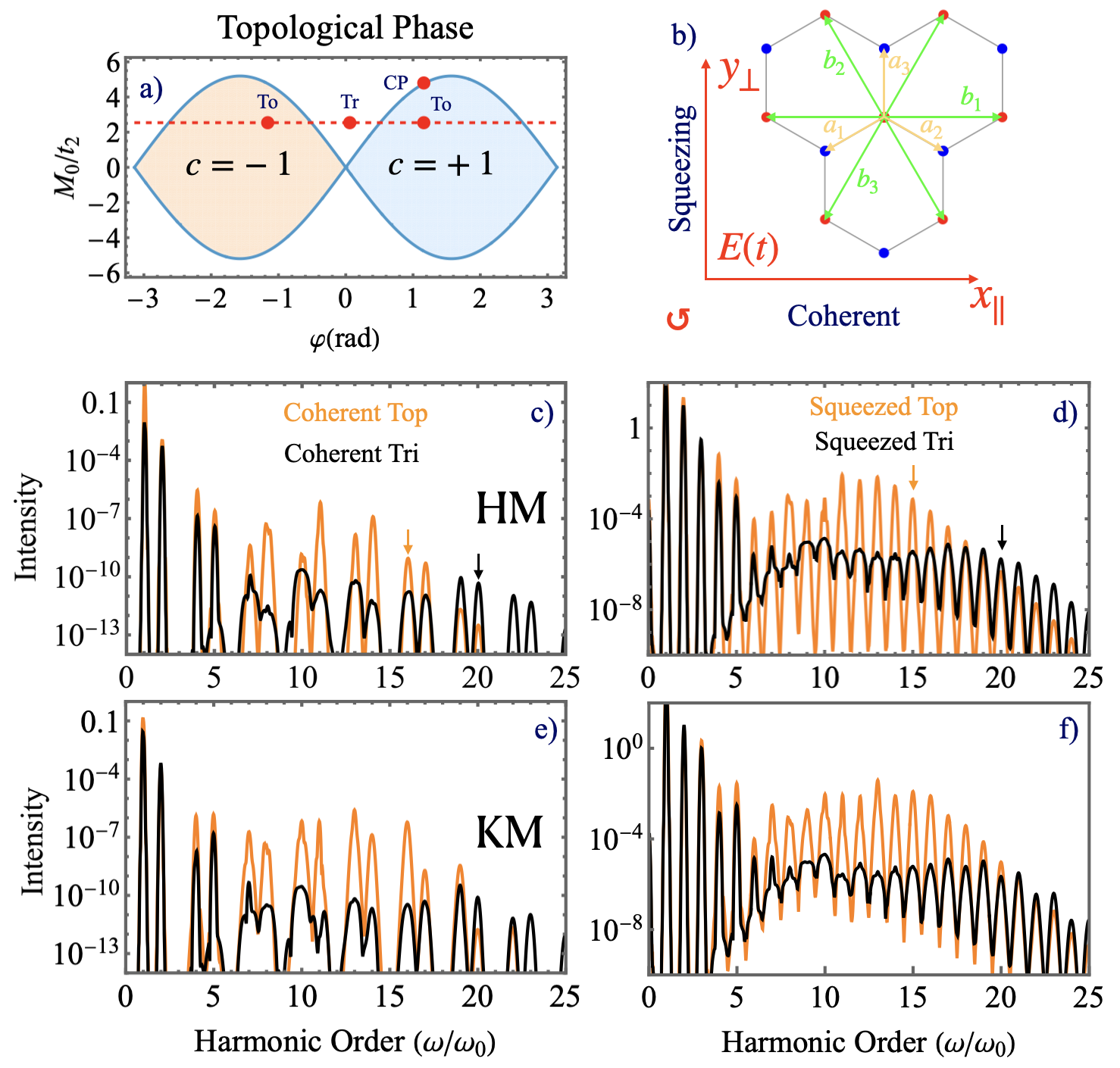}
\caption{Haldane model and harmonic spectra of the HM and KMM. (a) Phase diagram of the HM as a function of the magnetic flux $\varphi$ and the on-site potential $M_0$. The topological phases with Chern number $C=\pm1$ are separated from the trivial phase, $C=0$, by the phase boundary (blue line), at which the band gap closes. The red dashed line indicates $M_0/t_2=2.54$. The red points mark the parameters used in this work: the topological phases (``To'') at $\varphi=\pm1.16$~rad, with $C=\pm1$, the trivial phase (``Tr'') at $\varphi=0.06$~rad, and the critical point (``CP'') on the phase boundary. We used the hopping parameters $t_1=0.075$~a.u. and $t_2=t_1/3$, for which the trivial and topological phases have the same band gap, $\varepsilon_g\simeq3.0$~eV. (b) Crystal structure of the HM with the nearest-neighbor ($a_{1,2,3}$) and next-nearest-neighbor ($b_{1,2,3}$) vectors; the red and blue points represent the two sublattices. The coherent and squeezed components of the driving field are polarized along $x_\parallel$ and $y_\perp$, respectively. (c) Harmonic spectra of the HM driven by the coherent field in the topological (orange) and trivial (black) phases. (d) Same as (c) for the quantum driver. (e) and (f) Same as (c) and (d) for the KMM. The arrows mark the harmonic cutoff of each phase.}
\label{Fig1}
\end{figure}

\emph{Harmonic generation in Chern and topological insulators}---We start by introducing the response of the Chern insulator (see the Supplementary Material) to the circularly polarized quantum light. In particular, we will study the differences between the equal-time intensity correlation function, $g^{(2)}(0)$, calculated from the right circularly (RCP) and left circularly polarized (LCP) harmonics and how it differs for the topological and trivial phases. Most importantly, we will show that this result holds independently of the specific topological system considered here. In Fig.~\ref{Fig1}, we show the phase diagram of the HM in (a), together with the crystal structure in (b). The HHG spectra driven by a coherent field for the trivial (``Tri'', black line) and topological (``Top'', orange line) phases are presented in (c). The results obtained with the coherent driver agree with those reported in Ref.~\cite{Alexis1}. The spectra generated by the quantum driver are shown in panel (d) for the topological (orange line) and trivial (black line) phases.

Now, we turn our attention to the harmonic generation process for a TI, described by the KMM, and driven by the circularly polarized quantum driver. An example of the results comparing the classical and quantum driven HHG spectra is presented in Fig.~\ref{Fig1}~(e) and (f). The results likewise show a symmetry breaking for the quantum driver case and the increased harmonic yield for the topological matter.
However, a key question in high-harmonic spectroscopy of topological systems is whether the different topological systems exhibit a common response to light that could serve as a shared observable capable of distinguishing topological from trivial matter. For classical driving fields, this question was recently addressed in Ref.~\cite{Ofer1}, which found no evidence of a robust, material-independent topological signature in the HHG spectrum. More recent work, however, showed that channel-resolved intensity yields encode topological information through harmonic enhancements of up to two orders of magnitude relative to the trivial phase~\cite{Lorenzo_HHG_Topo}. More importantly, the detection of topological information in the TI is not possible via the circular dichroism since the total Chern number is zero, and needs the evaluation of the ellipticity-dependent harmonic yield~\cite{Kim_2023}.

Two important insights can be observed by comparing panels (c) and (d) in Fig.~\ref{Fig1}. First, the differences between the trivial and topological phases in both the harmonic yield and the harmonic cutoff are maintained regardless of the nature of the driver. Second, unlike the circularly polarized coherent driver, the quantum driver breaks the dynamical symmetry of the system, resulting in the generation of classically forbidden harmonics~\cite{stammer2026fluctuation}. Nevertheless, the quantum driver does not generate new features in the harmonic spectra that clearly distinguish the trivial and topological phases. Instead, the observed differences remain similar to those already obtained with the coherent driver, as reported in Refs.~\cite{Alexis1, Kim_2023}. However, the emerging quantum optical analysis of HHG allows one to access new observables, and we will show how such signatures lead to a share characterization of the two materials.

\begin{figure}[ht!]
\centering 
\includegraphics[width=1\linewidth,trim=0cm 0cm 0cm 0cm,clip]{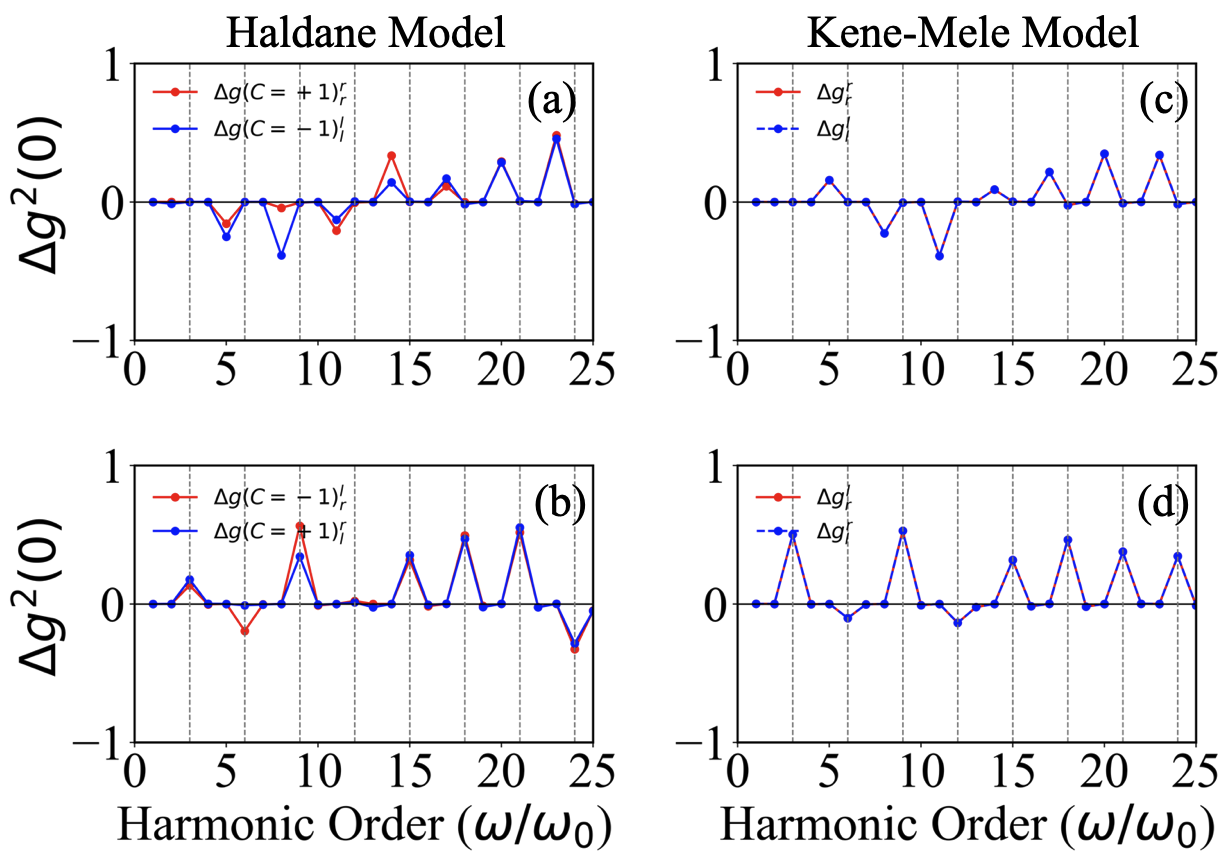}
\caption{Helical asymmetry of the $g^{(2)}(0)$ function. (a)--(d) Helical asymmetry $\Delta g^{i}_{j}$ of Eq.~\eqref{eq:deltag2} for the HM [(a) and (b)] and the KMM [(c) and (d)], with equal driver and harmonic helicities in (a) and (c) and opposite helicities in (b) and (d). Red and blue lines correspond to RCP and LCP harmonics, respectively, and for the HM the legends indicate the Chern number of the topological phase. (e) Circular-dichroic response $\mathcal{D}_C(m)$ for the topological phases of the HM with $C=\pm1$, the trivial phase and the critical point of the HM, and the topological phase of the KMM.}
\label{Topology}
\end{figure}

\emph{Symmetries in the $g^{(2)}(0)$ function}---While the harmonic intensity spectrum provides information about the average photon number, a measure of the underlying photon statistics is given by the second-order intensity correlation function $g^{(2)}(0)$~\cite{stammer2026photon, stammer2025theory}, defined as $g^{(2)}(0)=\langle \hat{b}^{\dagger 2}\hat{b}^{2}\rangle/\langle \hat{b}^{\dagger}\hat{b}\rangle^{2}$, where $\hat{b}$ ($\hat{b}^{\dagger}$) is the annihilation (creation) operator of the harmonic mode of a given order and helicity. In the following we will answer the crucial question: \emph{can the intensity correlation reveal differences between trivial and topological matter?} To answer this question, we compute $g^{(2)}(0)$ for four different points across the topological phase transition (see Fig.~\ref{Fig1}~(a)). These points correspond to two topological phases with different Chern numbers, $C=\pm1$, one point in the trivial phase, and one critical point separating the two phases. For a detailed discussion of the $g^{(2)}(0)$ functions for the RCP and LCP harmonics, we refer the reader to the Supplementary Material. Furthermore, we define the difference between the  $g^{(2)}(0)$ functions of the topological and trivial phases. This describes the asymmetry present in the $g^{(2)}(0)$ function calculated from the RCP and LCP harmonics for different phases
\begin{equation}\label{eq:deltag2}
    \Delta g^{i}_{j} = \frac{g^{(2),i}_{j,to}(0)-g^{(2),i}_{j,tr}(0)}{g^{(2),i}_{j,to}(0)+g^{(2),i}_{j,tr}(0)},
\end{equation}
here the upper index $i$ denotes the driver helicity, the lower index $j$ the emitted helicity, with $i,j\in\{R,L\}$, ``to'' stands for the topological phase and ``tr'' for the trivial phase. The helical asymmetry $\Delta g^{i}_{j}$ reveals the harmonic regions in which either phase dominates. For positive values of $\Delta g^{i}_{j}$, the dominant contribution originates from the topological matter, while for negative values it is the trivial phase that dominates. An example of the results for the different drivers and Chern numbers is presented in Fig.~\ref{Topology}. In all panels, red represents the RCP harmonics while blue represents the LCP ones, and for the HM the legends indicate the Chern number $C=\pm1$ of the topological phase entering Eq.~\eqref{eq:deltag2}. Interestingly, for the panels (a) to (d), the higher harmonics are dominated by the topological matter, a signature common to both types of matter. Moreover, $\Delta g^{i}_{j}$ in all the panels shows an excess of topological matter, for both models, in the harmonic spectra. In particular, when the driver polarization and the harmonic polarization are the same, the $\Delta g^{i}_{j}$ is dominated by the topological matter in the high harmonic region, Fig.~\ref{Topology}~(a) and (c). The dominance of the topological matter in the $\Delta g^{i}_{j}$ is more pronounced when the polarizations are opposite, Fig.~\ref{Topology}~(b) and (d).   

\emph{Discussion}---To gain further insight into the origin of the $\Delta g^{i}_{j}$ differences and the symmetries presented in Fig.~\ref{Topology}, we calculated analytically the $g^{(2)}(0)$ function for different responses of the harmonic emission to the fluctuations of the driving field (see the Supplementary Material). Under a coherent driving field, the harmonic yield is determined solely by the average nonlinear current, and the measured spectrum therefore reflects the mean electronic response of the material. For a quantum driver, the equal-time intensity correlation function $g^{(2)}(0)$ instead probes how the harmonic emission responds to the fluctuations of the driving field~\cite{stammer2026fluctuation}. Since the squeezing acts on a single quadrature, the field fluctuations are described by a real Gaussian variable $\xi$ with zero mean and variance $\sigma^{2}$ set by the squeezing parameter $r$. The amplitude of a given harmonic can then be expanded around the coherent field as $\mathcal{E}(\xi)\simeq c_{0}+c_{1}\xi+c_{2}\xi^{2}$, where $c_{0}$ is the amplitude generated by the coherent part of the driver, and the real coefficients $c_{1}$ and $c_{2}$ describe the linear and quadratic response of the harmonic to the fluctuation.

Within this description, $g^{(2)}(0)=\langle I^{2}\rangle/\langle I\rangle^{2}$, where $I=\mathcal{E}^{2}$ is the harmonic intensity and $\langle\cdot\rangle$ denotes the average over the Gaussian distribution of $\xi$. Two cases follow from the selection rules discussed above. For the classical harmonics ($3m-1$), $c_{0}\neq0$ and, keeping the leading linear term in $\xi$, $g^{(2)}(0)$ depends only on the ratio $w=c_{1}^{2}\sigma^{2}/c_{0}^{2}$ between the fluctuation-driven and the coherent emission. For the quantum harmonics ($3m$), which are forbidden for a circular coherent driver, $c_{0}=0$ and the emission is sourced entirely by the fluctuation, so that $g^{(2)}(0)$ depends only on the ratio $u=c_{2}^{2}\sigma^{2}/c_{1}^{2}$ between the quadratic and the linear response. The Gaussian average gives
\begin{equation}
    g^{(2)}(0)=
    \begin{cases}
        \dfrac{1+6w+3w^{2}}{(1+w)^{2}}, & (3m-1),\\[2ex]
        \dfrac{3+90u+105u^{2}}{(1+3u)^{2}}, & (3m).
    \end{cases}
    \label{eq:g2cases}
\end{equation}

For the classical harmonics, $g^{(2)}(0)$ increases monotonically from the coherent value $g^{(2)}(0)=1$ for $w\to0$ to $g^{(2)}(0)=3$ for $w\to\infty$. For the quantum harmonics, $g^{(2)}(0)=3$ when the response to the fluctuation is linear ($u=0$)~\cite{Javi1}, while any quadratic contribution ($u>0$) gives $g^{(2)}(0)>3$, with $g^{(2)}(0)\to35/3$ for $u\to\infty$. A value $g^{(2)}(0)=3$ therefore signals a harmonic response that is linear in the field fluctuation, even though the harmonic emission is a highly nonlinear function of the driving field, whereas $g^{(2)}(0)>3$ requires a nonlinear response to the fluctuation.

Following Eq.~\eqref{eq:symmetries} (see the End Matter), $\Delta g^{i}_{j}$ is nonzero only at the classical $3m-1$ harmonics when the driver and harmonic helicities coincide [Fig.~\ref{Topology}(a) and (c)], and only at the quantum $3m$ harmonics when the helicities are opposite [Fig.~\ref{Topology}(b) and (d)]. A positive value of $\Delta g^{i}_{j}$ means that $g^{(2)}(0)$ is larger in the topological phase than in the trivial phase. In both models, positive values of $\Delta g^{i}_{j}$ dominate the high-order harmonics near the cutoff.

Our hypothesis on the microscopic origin of these differences can be read as follows. The quantum driver generates a distribution of Bloch trajectories $\mathbf{k}_{\alpha}(t)=\mathbf{k}_{\rm cl}(t)+\delta\mathbf{k}_{\alpha}(t)$, where $\mathbf{k}_{\rm cl}(t)$ is the classical trajectory and $\delta\mathbf{k}_{\alpha}(t)$ the deviation induced by the field fluctuation. This distribution modifies the electronic response beyond the classical trajectory and introduces corrections to the quantities governing HHG, including the band velocity, Berry curvature, and interband polarization, which set the coefficients $c_{1}$ and $c_{2}$. The magnitude of these corrections is determined by the momentum-space structure of the bands, in particular by the Berry-curvature distribution and the regions of the Brillouin zone sampled by the driven electron. As a consequence, the ratios $w$ and $u$, and with them $g^{(2)}(0)$, differ between the topological and trivial phases. Since $g^{(2)}(0)$ probes intensity fluctuations rather than the average harmonic yield, it reveals these differences more clearly, particularly for harmonics near the cutoff where the nonlinear response is enhanced. Further analysis and numerical calculations are necessary to confirm this physical picture.

\emph{Conclusions}---We have introduced the quantum-optical quantity, the helical intensity correlation function $\Delta g^{i}_{j}$, which compares the helicity-resolved $g^{(2)}(0)$ functions of high harmonics driven by quantum light in the topological and trivial phases. Applied to a Chern insulator, described by the HM, and to a time-reversal-invariant $\mathbb{Z}_2$ topological insulator, described by the KMM, it distinguishes the two phases in the same way: for the parameters considered, $\Delta g^{i}_{j}$ is predominantly positive for the high-order harmonics near the cutoff, that is, $g^{(2)}(0)$ is larger in the topological than in the trivial phase. This behavior is common to the two systems, whereas the classical observables of the HHG spectrum that reveal their topology differ from one system to the other~\cite{Alexis1,Kim_2023,Ofer1}. In the HM, the circular-dichroic response $\mathcal{D}_C(m)$ reverses sign with the Chern number, while in the KMM it vanishes, as required by time-reversal symmetry. Within a Gaussian description of the field fluctuations, these differences are consistent with a distinct response of the harmonic emission to the quantum fluctuations of the driver in the two phases. Since HHG driven by bright squeezed vacuum has been demonstrated in solids~\cite{rasputnyi_HHG, PB2025} and $g^{(2)}(0)$ of a single harmonic and helicity can be obtained from intensity-correlation measurements, harmonic photon correlations offer a route to probe topological matter with quantum light.

\begin{acknowledgments}

\emph{Acknowledgments}---ICFO-QOT group acknowledges support from:
European Research Council AdG NOQIA;
MCIN/AEI (PGC2018-0910.13039/501100011033, CEX2019-000910-S/10.13039/501100011033, Plan National FIDEUA PID2019-106901GB-I00, Plan National STAMEENA PID2022-139099NB, I00, project funded by MCIN/AEI/10.13039/501100011033 and by the “European Union NextGenerationEU/PRTR" (PRTR-C17.I1), FPI); QUANTERA DYNAMITE PCI2022-132919, QuantERA II Programme co-funded by European Union’s Horizon 2020 program under Grant Agreement No 101017733; Ministry for Digital Transformation and of Civil Service of the Spanish Government through the QUANTUM ENIA project call - Quantum Spain project, and by the European Union through the Recovery, Transformation and Resilience Plan - NextGenerationEU within the framework of the Digital Spain 2026 Agenda;
Fundació Cellex;
Fundació Mir-Puig;
Generalitat de Catalunya (European Social Fund FEDER and CERCA program);
Barcelona Supercomputing Center MareNostrum (FI-2023-3-0024);
Funded by the European Union. Views and opinions expressed are however those of the author(s) only and do not necessarily reflect those of the European Union, European Commission, European Climate, Infrastructure and Environment Executive Agency (CINEA), or any other granting authority. Neither the European Union nor any granting authority can be held responsible for them (HORIZON-CL4-2022-QUANTUM-02-SGA PASQuanS2.1, 101113690, EU Horizon 2020 FET-OPEN OPTOlogic, Grant No 899794, QU-ATTO, 101168628), EU Horizon Europe Program (This project has received funding from the European Union’s Horizon Europe research and innovation program under grant agreement No 101080086 NeQST Grant Agreement 101080086 — NeQST);
ICFO Internal “QuantumGaudi” project. J.~R.-D.~acknowledges funding from UKRI2300 -- Attosecond Photoelectron Imaging with Quantum Light. A.C. acknowledges the Sistema Nacional de Investigaci\'on de Panam\'a for partial financial support. DK thanks the Institute for Basic Science (IBS), Korea under Project Code IBS-R014-A1. The National Research Foundation of Korea grants (Grant no. NRF-2023R1A2C2007998). The MSIT (Ministry of Science and ICT), Korea, under the ITRC (Information Technology Research Center) support program (Grant no. IITP-2023-RS-2022-00164799) supervised by the IITP (Institute for Information \& Communications Technology Planning \& Evaluation). W. Gao thanks financial support from National Natural Science Foundation of China (Grant number: 12474309).

\end{acknowledgments}

\bibliography{apssamp}

\clearpage
\section{Appendix A}

Further analysis reveals that a symmetry in the $\Delta g^{i}_{j}$ emerges consistent with the symmetries identified in Ref.~\cite{QL_Solids} for a perfectly circular quantum driver, Fig.~\ref{Topology}~(a) to (d): The $3m$ harmonics, with $m$ a positive integer, obey an equivalence between the R(L) and L(R) channels of the laser and harmonic helicity, respectively. Similarly, the $3m-1$ harmonics obey an equivalence between the R(R) and L(L) channels
\begin{eqnarray}
    \text{for }  (3m-1) &:& \Delta g^R_R \longleftrightarrow \Delta g^L_L, \nonumber \\
   \text{for }  (3m) &:& \Delta g^R_L \longleftrightarrow \Delta g^L_R. \label{eq:symmetries}
\end{eqnarray}

These two channels are precisely the classical and quantum harmonics distinguished by the symmetry-breaking of the selection rules for perfect circularly polarized light. These symmetry rules (see the Supplementary Material) allow us to define a circular-dichroic $g^{(2)}$ response
\begin{equation}
    \mathcal{D}_C(m) = \mathcal{S}_C^R(m) - \mathcal{S}_C^L(m),
\end{equation}
where $\mathcal{S}_C^{R}(m)=g^{(2),R}_{R,C}(0)\big|_{3m-1}+g^{(2),R}_{L,C}(0)\big|_{3m}$ and $\mathcal{S}_C^{L}(m)=g^{(2),L}_{L,C}(0)\big|_{3m-1}+g^{(2),L}_{R,C}(0)\big|_{3m}$ sum the classical and quantum channels for each driver helicity. Here the label $C$ indicates the phase with Chern number $C$, and the subscript of the vertical bar gives the harmonic order. The response $\mathcal{D}_C(m)$ changes sign under a reversal of the Chern number
\begin{equation}
    \mathcal{D}_{-C}(m) \simeq -\mathcal{D}_C(m).  \label{eq:DChern}
\end{equation}

\begin{figure}[h!]
\centering 
\includegraphics[width=1\linewidth,trim=.6cm 1cm 0.6cm 0cm,clip]{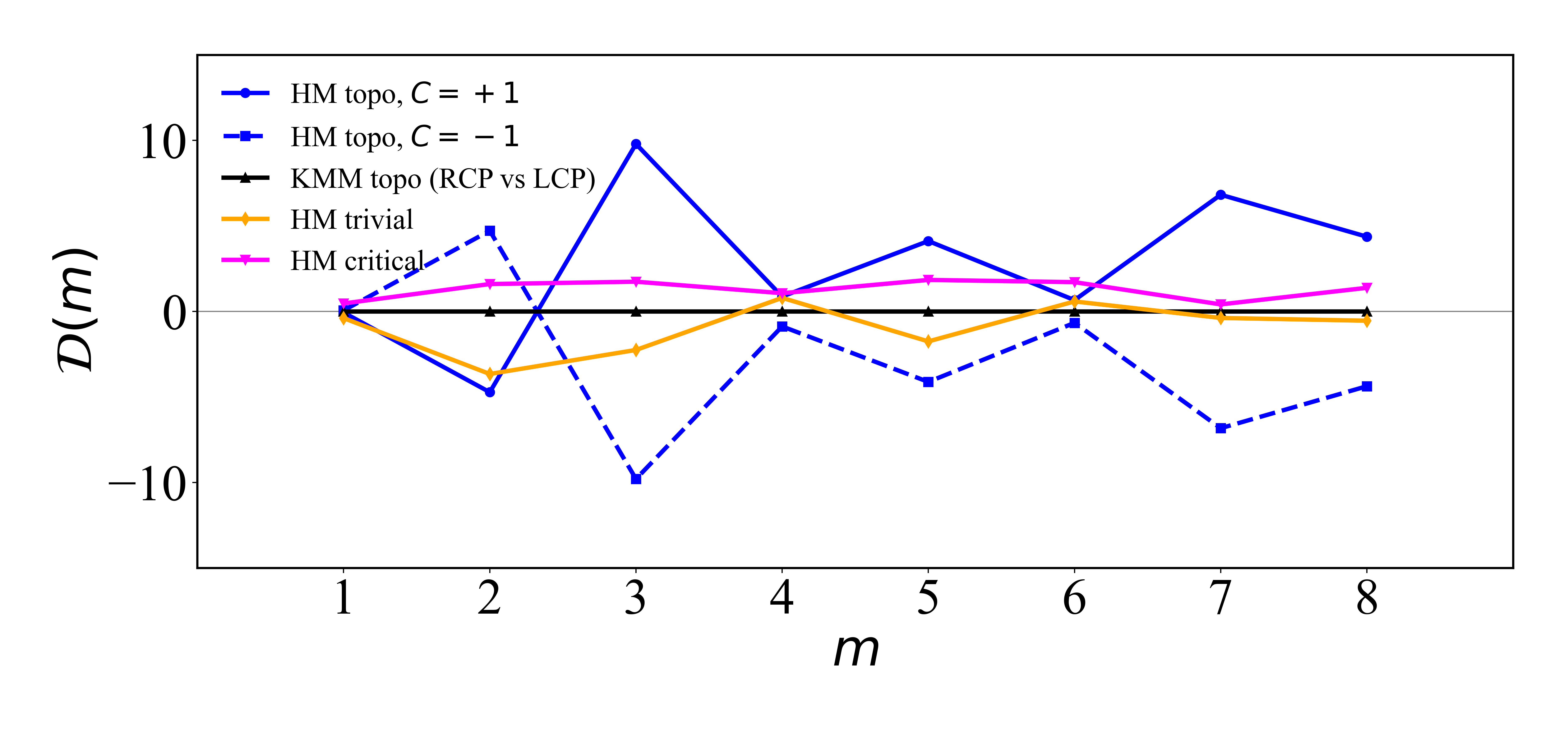}
\caption{Circular-dichroic response $\mathcal{D}_C(m)$ for the topological phases of the HM with $C=\pm1$, the trivial phase and the critical point of the HM, and the topological phase of the KMM. For calculating $\mathcal{D}_C(m)$ we used the same $g^{(2)}(0)$ values used to generate Fig.~\ref{Topology}.}
\label{Topology2}
\end{figure}

Figure~\ref{Topology2} shows $\mathcal{D}_C(m)$ for the two topological phases of the HM, together with the same quantity evaluated for the trivial phase and the critical point of the HM and for the topological phase of the KMM, for which it vanishes at all orders. For the HM, the sign of $\mathcal{D}_C(m)$ varies with $m$, but its sum over all harmonics is positive for $C=+1$ and negative for $C=-1$. A normalized circular-dichroism-type quantity $\bar{\mathcal{D}}_C$ (see the Supplementary Material) condenses this behavior into a single number, $\bar{\mathcal{D}}_{C=+1}=+0.2459$ and $\bar{\mathcal{D}}_{C=-1}=-0.2459$. The magnitude of the Chern number can be calculated from the circular dichroism of the harmonics resulting from the quantum driver (see the Supplementary Material), in the same way as for the classical counterpart~\cite{Alexis1}. 

\clearpage
\onecolumngrid

\setcounter{section}{0}
\setcounter{equation}{0}
\setcounter{figure}{0}
\setcounter{table}{0}
\renewcommand{\thesection}{S\Roman{section}}
\renewcommand{\theequation}{S\arabic{equation}}
\renewcommand{\thefigure}{S\arabic{figure}}
\renewcommand{\thetable}{S\arabic{table}}
\renewcommand{\theHsection}{S\Roman{section}}
\renewcommand{\theHequation}{S\arabic{equation}}
\renewcommand{\theHfigure}{S\arabic{figure}}
\renewcommand{\theHtable}{S\arabic{table}}

\begin{center}
{\large\bfseries Supplementary Material for ``Attosecond Quantum Optical Spectroscopy of Topological Materials''}\\[2ex]
C.~Granados, J.~Rivera-Dean, D.~Kim, A.~Chac\'on, W.~Gao, and P.~Stammer
\end{center}
\vspace{2ex}

\section{High harmonic generation in solids}

To calculate the high-harmonic generation in the solid, we solve the time-dependent density matrix equation (TDDM) \cite{KimSym2022, DasolKMM}. To avoid the computational problems stemming from the singularity of the Bloch wave functions for topological materials \cite{Alexis1, Silva1}, we solve the TDDM in the Wannier gauge. This gauge is constructed from localized Wannier functions. Therefore, the electron dynamics under the action of the external electric field is described by the reduced density matrix $\rho_{mn}^{(W)}$: 
\begin{equation}\label{eq:DenMat}
    i\frac{\partial }{\partial t}\rho_{mn}^{(W)}(\mathbf{K},t)=\Big[H_0^{(W)}(\mathbf{k}(t)),\rho^{(W)}(\mathbf{K},t) \Big]_{mn}
\end{equation}
where $\mathbf{k}(t)=\mathbf{K}+\mathbf{A}(t)$, $H_0^{(W)}(\mathbf{k}(t))$ is the unperturbed Hamiltonian. The sub-indices $mn$, represent the different bands, which for our case are the valence band and the conduction band. The generated current can be calculated as 
\begin{equation}\label{eq:TotalCurrent}
\mathbf J(t) = \sum_{m,n}\int_{BZ} d\mathbf{K} \mathbf{P}_{mn}^\text{(W)}(\mathbf{K} + \mathbf{A}(t)) \rho_{mn}^{(\text{W})}(\mathbf{K},t),
\end{equation}
where, $\mathbf{P}_{mn}^W(\mathbf{k}) $, is the momentum matrix element in the Wannier gauge, and can be calculated as 
\begin{equation}
\mathbf{P}_{mn}^\text{(W)}(\mathbf{k})  = \langle \phi^{\text{(W)}}_{m\mathbf k} |\partial \hat{H}_0^{\text{(W)}}(\mathbf k)| \phi^{\text{(W)}}_{m\mathbf k} \rangle
\end{equation}
The total current, once again can be calcualted via interband  
\begin{equation}
\mathbf J_{re}(t) = \sum_{m\ne n}\int_{BZ} d\mathbf{K} \mathbf{P}_{mn}^\text{(H)}(\mathbf{K} + \mathbf{A}(t)) \rho_{mn}^{\text{(H)}}(\mathbf{K},t),
\end{equation}
and intraband contributions 
\begin{equation}
\mathbf J_{ra}(t) = \sum_{m}\int_{BZ} d\mathbf{K} \mathbf{P}_{mm}^\text{(H)}(\mathbf{K} + \mathbf{A}(t)) \rho_{mm}^{\text{(H)}}(\mathbf{K},t),
\end{equation}
Here, the index “H" refers to the Hamiltonian Gauge. The relationship between both the Hamiltonian and Wannier is customary and can be found in Ref.~\cite{KimSym2022}, for example. The total harmonic spectra, $\mathbf J(t)=\mathbf J_{intra}(t)+\mathbf J_{inter}(t)$ is calculated from the Fourier transform of the individual inter and intraband contributions. 
\begin{equation}
    I_{HHG}(\omega)=\Bigg| \mathcal{F}\Big[ J_{intra}(t)+J_{inter}(t)\Big] \Bigg|^2.
\end{equation} 
Because we are interested in the effect of the quantum light on the topological matter, for the solid we used two distinct models, the Haldane model (HM) and the Kane-Mele model (KMM). The two models represent two different classes of topological materials: (i) a Chern insulator and (ii) a Topological insulator. In particular, the HM describes the phase transition from a trivial to a topological insulator by introducing a magnetic flux, $\varphi$, and a complex next-nearest-neighbor (NNN) hopping parameter, $t_2$, to the graphene Hamiltonian. The Hamiltonian of the system is given by \cite{bernevig2013topological}: 
\begin{equation}
H_{T} = -t_1\sum_{\langle i,j\rangle}c_i^\dagger c_j  - t_2 \sum_{\langle \langle i,j\rangle\rangle} e^{i\varphi_{i,j}} c_i^\dagger c_j + M\sum_i \epsilon_i  c_i^\dagger c_i.
\end{equation}
Here, the operators $c_i^\dagger/c_j$ creates/annihilates a (spinless) fermion at the site $i$ and $\epsilon_i  = \pm 1$, depending on the sub-lattice $A$ or $B$.  The nearest-neighbor-hopping is represented by $t_1$, and the on-site inversion symmetry-breaking mass term by $M$. Most importantly, the trivial and topological matter described by the HM, can be characterized by a topological invariant, the Chern number, $C$. For topological matter $C=\pm 1$ and for trivial matter $C=0$. 

The KMM, represents a topological insulator characterized by a different topological invariant, the Z$_2$ \cite{KMM}. It is an extension of the HM to include the spin-orbit coupling via the NNN hopping parameter $t_2\sin(\varphi)$. It also includes the Rashba term, describing an external field, perpendicular to the current in the solid \cite{KMM}. The Hamiltonian of the model, is often described as two HM with opposite signs of the complex NNN hopping parameter, and reads as follows 
\begin{eqnarray}
H_{\rm KM} &=& -t_1\sum_{\langle i,j\rangle,\sigma}
c_{i\sigma}^{\dagger}c_{j\sigma} +i t_{2}\sum_{\langle\!\langle i,j\rangle\!\rangle,\sigma,\sigma'}
\nu_{ij} c_{i\sigma}^{\dagger} (s_z)_{\sigma\sigma'}
c_{j\sigma'} \nonumber \\
&+& \lambda_{SO} \sum_{i,\sigma} \epsilon_i c_{i\sigma}^{\dagger}c_{i\sigma}.
\end{eqnarray}
Importantly and differently to the HM, the KMM respects the time reversal symmetry. 

\section{Equal time intensity correlation function calculations}

Here we extend the equal-time correlation function $g^{(2)}(0)$ of the main text to arbitrary order $g^{(n)}(0)$. 

\subsection{Gaussian sampling of the harmonic intensity}
\label{sec:SM_sampling}

In the classical-limit (Husimi-marginal) picture, expectation values of the emitted radiation are obtained by averaging the semiclassical harmonic response over the field fluctuation. Because the squeezing acts on a single quadrature, the relevant fluctuation is the real Gaussian variable $\xi$ with
\begin{equation}
\langle\xi\rangle = 0,\qquad \langle\xi^2\rangle=\sigma^2,\qquad
\langle\xi^{2k}\rangle=(2k-1)!!\sigma^{2k},\qquad \langle\xi^{2k+1}\rangle=0 .
\label{eq:SM_moments}
\end{equation}
We can then write
$\langle \cdot \rangle=\int d^2\alpha Q(\alpha)(\cdot)$ as a 
one-dimensional Gaussian integral
\begin{equation}
\langle I_q^{n}\rangle
= \int_{-\infty}^{\infty}\frac{d\xi}{\sqrt{2\pi\sigma^2}}
e^{-\xi^2/2\sigma^2}A_q(\xi)^{2n},
\qquad
A_q=c_0+c_1\xi+c_2\xi^2,\qquad I_q=A_q^2 ,
\label{eq:SM_integral}
\end{equation}
with $c_n\equiv J_q^{(n)}$ the response coefficients evaluated along the classical trajectory and projected onto $\omega_q$. The coefficients are taken real for clarity. The normalized correlations are
\begin{equation}
g^{(n)}(0)=\frac{\langle I_q^{n}\rangle}{\langle I_q\rangle^{n}} .
\label{eq:SM_gn_def}
\end{equation}
Throughout we use the double-factorial identity
\begin{equation}
\binom{2n}{2j}(2j-1)!!=\frac{(2n)!}{(2n-2j)!2^{j}j!}.
\label{eq:SM_identity}
\end{equation}

\subsection{Case one: $c_0\neq0$}

For the classically allowed orders we keep the dominant linear fluctuation,
$A_q=c_0+c_1\xi$. Expanding the $2n$-th power and discarding the odd moments
of Eq.~\eqref{eq:SM_moments}, only even powers $\xi^{2j}$ survive, so that the
binomial sum $\sum_{k=0}^{2n}$ reduces to $\sum_{j=0}^{n}$ under $k=2j$:
\begin{align}
\langle I_q^{n}\rangle
&=\big\langle (c_0+c_1\xi)^{2n}\big\rangle
=\sum_{j=0}^{n}\binom{2n}{2j}c_0^{2n-2j}c_1^{2j}(2j-1)!!\sigma^{2j}
\nonumber\\
&=c_0^{2n}\sum_{j=0}^{n}\binom{2n}{2j}(2j-1)!!w^{j}
\;\equiv\; c_0^{2n}H_n(w),
\qquad
w\equiv\frac{c_1^2\sigma^2}{c_0^2}.
\label{eq:SM_allowed_moment}
\end{align}
Using Eq.~\eqref{eq:SM_identity},
\begin{equation}
H_n(w)=\sum_{j=0}^{n}\binom{2n}{2j}(2j-1)!! w^{j}
=(2n)!\sum_{j=0}^{n}\frac{w^{j}}{(2n-2j)!2^{j}j!}.
\label{eq:SM_Hn}
\end{equation}
The important prefactor $c_0^{2n}$ cancels, leaving
\begin{equation}
g^{(n)}(0)=\frac{H_n(w)}{(1+w)^{n}}
\label{eq:SM_gn_allowed}
\end{equation}
a function of the fluctuation-to-coherent ratio $w$ alone. The first members are
$H_1=1+w$ and $H_2=1+6w+3w^2$, recovering
$g^{(2)}=(1+6w+3w^2)/(1+w)^2$. Here for the limit $w\rightarrow 0$ we recover the coherent value $g^{(2)}=1$. For $w \rightarrow  \infty$ we recover $g^{(2)}=3$. Here $w \rightarrow  \infty$ means that $c_0$ is not important. 

\subsection{Case 2: $c_0=0$}

For the symmetry-forbidden orders the emission is sourced entirely by the
fluctuation, $A_q=c_1\xi+c_2\xi^2$, so that
$I_q^{n}=(c_1\xi+c_2\xi^2)^{2n}$ carries an overall factor $\xi^{2n}$. Keeping the even-power terms
\begin{align}
\langle I_q^{n}\rangle
&=\big\langle (c_1\xi+c_2\xi^2)^{2n}\big\rangle
=\sum_{j=0}^{n}\binom{2n}{2j}c_1^{2n-2j}c_2^{2j}
\langle\xi^{2n+2j}\rangle
\nonumber\\
&=\sum_{j=0}^{n}\binom{2n}{2j}c_1^{2n-2j}c_2^{2j}(2n+2j-1)!!\sigma^{2n+2j}
\nonumber\\
&=(c_1^2\sigma^2)^{n}\sum_{j=0}^{n}\binom{2n}{2j}(2n+2j-1)!!u^{j}
\;\equiv\;(c_1^2\sigma^2)^{n}G_n(u),
\qquad
u\equiv\frac{c_2^2\sigma^2}{c_1^2}.
\label{eq:SM_forbidden_moment}
\end{align}
The extra $\xi^{2n}$ shifts every Gaussian moment from $\langle\xi^{2j}\rangle$
to $\langle\xi^{2n+2j}\rangle$, i.e. it shifts the double-factorial argument
$(2j-1)!!\to(2n+2j-1)!!$ relative to Eq.~\eqref{eq:SM_Hn}. With
$\langle I_q\rangle=c_1^2\sigma^2G_1(u)=c_1^2\sigma^2(1+3u)$,
\begin{equation}
g^{(n)}(0)=\frac{G_n(u)}{(1+3u)^{n}},\qquad
G_n(u)=\sum_{j=0}^{n}\binom{2n}{2j}(2n+2j-1)!!u^{j}
\label{eq:SM_gn_forbidden}
\end{equation}
The first members are $G_1=1+3u$ and $G_2=3+90u+105u^2$, recovering the main-text result $g^{(2)}=(3+90u+105u^2)/(1+6u+9u^2)$. Here once again we can make $u \rightarrow  0$ we recover $g^{(2)}=3$. For $u \rightarrow  \infty$ we recover $g^{(2)}=11.7$. 

\section{Equal time intensity correlation function}

For the topological phases with , for a right-circularly polarized (RCP) driver, the results are presented in Fig.~\ref{Fig2_Alt}~(a1) and (c1), respectively. 
The corresponding results for a left-circularly polarized (LCP) driver are presented in panels (b1) and (d1). The results for the trivial phase are presented in panels (e1) and (f1), while those for the critical point are shown in panels (g1) and (h1). There are clear differences in the $g^{(2)}(0)$ functions among the different phases, particularly for the higher-order harmonics, for which the intensity correlation function grows beyond the previously reported value, i. e. $g^{2}(0) = 3 $ for atoms~\cite{stammer2026fluctuation} and solids~\cite{QL_Solids}. In all the panels, the red lines and blue lines represent the RCP and LCP harmonic polarization component, respectively.  

\begin{figure*}[ht!]
\centering 
\includegraphics[width=1\linewidth,trim=8cm 12cm 8cm 0cm,clip]{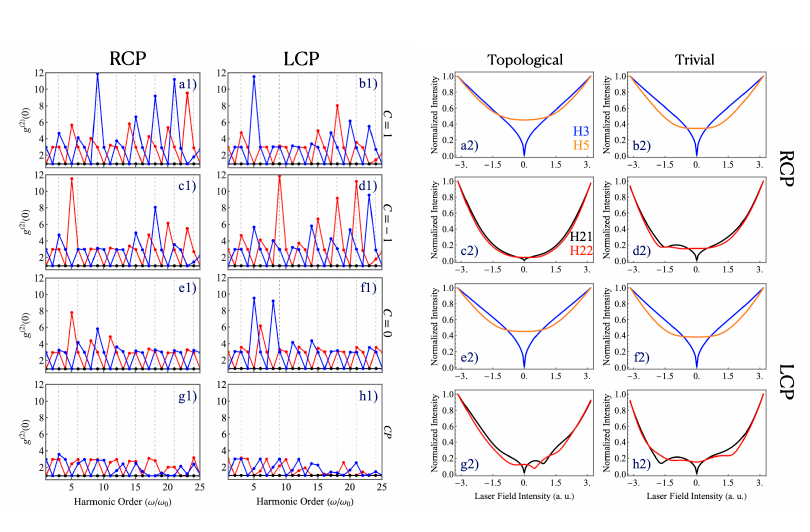}
\caption{Equal time intensity correlation function, $g^{(2)}(0)$, and normalized harmonic intensity for the HM. In panels (a1) and (b1) $g^{(2)}(0)$ function for RCP and LCP drivers and a topological phase with Chern number $C=+1$. In panels (c1) and (d1) $g^{(2)}(0)$ function the same drivers but for a topological phase with Chern number $C=-1$. In panels (g1) and (h1) we show the $g^{(2)}(0)$ function for the critical point. We note that for a determined topological phase, the quantum allowed harmonics polarization contribution corresponds to that dominant in the harmonic spectra. When the quantum driver field polarization does not align with the sign of the Chern number, the $g^{(2)}(0)$ function is not very sensitive to the Chern number, as shown in panels (b1) and (c1). If the polarization and Chern number align, then the quantum allowed harmonics polarization is the same as the Chern number. In all the panels, the red lines and blue lines represent the RCP and LCP harmonic polarization component. In panels (a2) to (d2) we show the 3$^\text{rd}$, 5$^\text{th}$, 21$^\text{st}$ and 22$^\text{nd}$ harmonic scaling for topological and trivial phases and for a RCP driving field. In panels (e2) to (h2), the corresponding results for a LCP driving field. The different scaling laws are the origin of the different $g^{(2)}(0)$ functions for the different phases. For constructing the $g^{(2)}(0)$ function plots, we used a field intensity $1.0\times10^{-9}$~a. u. for the quantum component.}
\label{Fig1_Alt}
\end{figure*}

\begin{figure*}[ht]
\centering 
\includegraphics[width=1\linewidth,trim=8cm 12cm 8cm 0cm,clip]{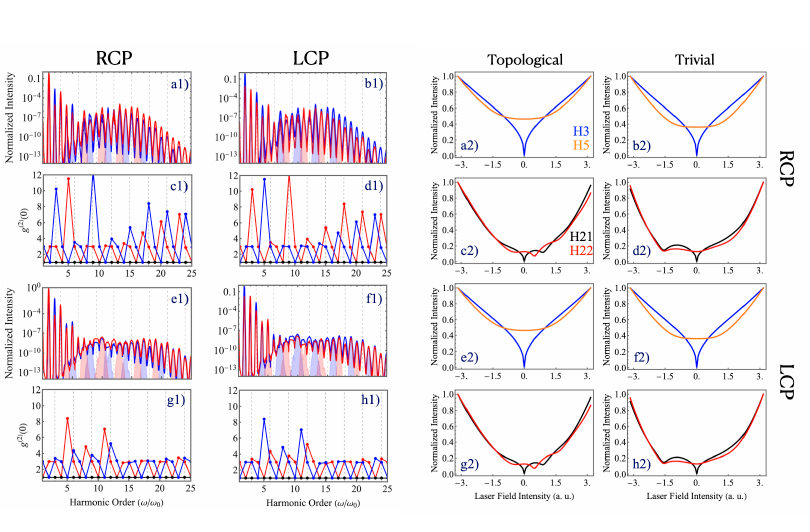}
\caption{High harmonic generation and normalized harmonic intensity calculated for the KMM. In panels (a1) and (b1) show the resulting harmonic spectra for a RCP and LCP drivier and for the topological phase, respectively. The resulting helical components are shown in red (RCP) and blue (LCP). The results for the trivial phase are presented in the panels (g1) and (h1). The $g^{(2)}(0)$ function is presented in panels (c1) and (d1) for the topological phase and (g1) and (h1) for the trivial phase. In panels (a2) to (d2) we show the 3$^\text{rd}$, 5$^\text{th}$, 21$^\text{st}$ and 22$^\text{nd}$ harmonic scaling for topological and trivial phases and for a RCP driving field. In panels (e2) to (h2), the corresponding results for a LCP driving field. We note that for constructing the $g^{(2)}(0)$ function plots, we used an intensity in the quantum component of $1.0\times10^{-9}$~a. u. }
\label{Fig2_Alt}
\end{figure*}

The results presented in Fig.~\ref{Fig1_Alt} (left panel) show a clear effect of the driver polarization on the resulting $g^{(2)}(0)$ function. Classically, the harmonics obey the selection rules $q=3n \pm 1$, with $n=1,2,3,\ldots$. For an RCP driver, the $q=3n-1$ harmonics have the same helicity as the driving field, whereas the $q=3n+1$ harmonics have the opposite helicity. For an LCP driver, the harmonic helicities are reversed. However, the quantum driver changes this picture. The resulting harmonics in all phases exhibit mixed polarization. The classically allowed harmonics contain a polarization component with Poissonian statistics ($g^{2}(0)=1$). In contrast, the newly generated harmonics, which arise from dynamical symmetry breaking, exhibit super-Poissonian statistics with $g^{2}(0)>1$. The main differences in the photon statistics appear in the high-harmonic region, where $g^{2}(0)>3$ for the topological phases. In contrast, for the trivial phase, $g^{2}(0)=3$ for the same harmonic orders. A similar behavior is observed at the critical point. Additionally, the lower-order harmonics at the critical point also retain the value $g^{2}(0)=3$. Consequently, the intensity correlation allows us to differentiate the trivial from the topological phase as well as from the phase transition point, i.e. the critical point. 

The origin of the different photon statistics can be correlated with the matter response to the driving field. In Fig.~\ref{Fig1_Alt}~(a2) to (h2), we present the harmonic yield as a function of the driving-field intensity for the harmonics 3$^\text{rd}$, 5$^\text{th}$, 21$^\text{st}$, and 22$^\text{nd}$, and for both the trivial and topological phases. The harmonics orders were selected to cover both the low and high harmonic regions as well as the classically allow and forbidden harmonics. Similar results were found for the low-order harmonics in both phases, where the photon statistics are also similar. However, in the high-harmonic region, the harmonic intensity differs significantly between the two phases. 

The results presented in Fig.~\ref{Fig1_Alt}~(a1) to (h1) are also consistent with the the harmonic yield dependence to the helicity of the driver. The harmonic yield exhibits distinct responses for LCP and RCP light when the harmonic process is driven in a topological phase. This helicity-dependent behavior was exploited in Ref.~\cite{Alexis1} to demonstrate that the circular dichroism of the high-harmonic emission maps the Chern number across the different topological phases of the Haldane model. Similar results were obtain by comparing the helicity of the individual harmonics~\cite{Silva1} for the trivial and topological phases of the HM.

Figure~\ref{Fig2_Alt}(a1) and (b1) show examples of the harmonic spectra for the topological phase, while panels (e1) and (f1) display the corresponding spectra for the trivial phase. The associated equal-time intensity correlation functions, $g^{(2)}(0)$, are presented in panels (c1) and (d1) for the topological phase and in panels (g1) and (h1) for the trivial phase. Consistent with the coherent case, the harmonic yield is larger in the topological phase than in the trivial phase~\cite{Alexis1,Kim_2023}. However, under quantum driving, the squeezed field breaks the classical dynamical symmetry, giving rise to distinct harmonic spectra and revealing purely quantum features that are absent for a coherent driver.

Additionally, the calculated $g^{(2)}(0)$ functions for the topological and trivial phases exhibit the same qualitative behavior observed for high-harmonic generation in the Haldane model. While the trivial phase retains the characteristic value $g^{(2)}(0)=3$ for most harmonic orders, the topological phase exhibits a pronounced enhancement of the intensity correlation function, with $g^{(2)}(0)>3$. This clear contrast between the two phases demonstrates that the equal-time intensity correlation function provides a robust observable for distinguishing topological from trivial phases under quantum-light driving.
Similar to the HM, we examine the harmonic yield as a function of the driving field intensity for the KMM, as shown in Fig.~\ref{Fig2_Alt} (a2) to (h2). The results show clear differences for the harmonic response in the higher harmonics. The results point to the nonlinear response of the material as the microscopic mechanism explaining the differences observed in the $g^{(2)}(0)$ functions. 

\section{Topological signatures}

The goal is to identify signatures in the photon statistics that are common to all models and reflect the topology of the underlying phase. To this end, we first seek a measure that discriminates the trivial phase ($C=0$) from the topological phases ($|C|=1$), and subsequently one that additionally distinguishes $C=+1$ from $C=-1$.

The magnitudes of the individual harmonics do not provide a clear distinction between the topological and the trivial phase. However, the differences $\Delta g$ exhibit an underlying symmetry that mirrors the one known from solids~\cite{QL_Solids}. Here, the $3m-1$ harmonics show equivalent behavior for the R(R) and L(L) channels, while the $3m$ harmonics show equivalent behavior for the R(L) and L(R) channels, where the first helicity denotes the drive and the helicity in brackets the emitted one. With the upper index denoting the drive helicity, the lower index the emitted helicity, and $m$ an integer, this equivalence reads
\begin{align}
    \Delta g^R_R(3m-1) & \longleftrightarrow \Delta g^L_L(3m-1), \\
    \Delta g^R_L(3m) & \longleftrightarrow \Delta g^L_R(3m).
\end{align}
These two families correspond, respectively, to the classically allowed ($3m-1$) and the classically forbidden, or quantum ($3m$), harmonics of the symmetry-breaking argument.

To exploit this symmetry without relying on individual harmonics, whose magnitudes and signs can differ considerably, we group the symmetry-equivalent channels and average over each pair. This separates the classically allowed from the classically forbidden contributions,
\begin{align}
    G_{\mathrm{cl}}(m) & \equiv \frac{1}{2} \left[ \Delta g^R_R(3m-1) + \Delta g^L_L(3m-1) \right], \\
    G_{\mathrm{q}}(m) & \equiv \frac{1}{2} \left[ \Delta g^R_L(3m) + \Delta g^L_R(3m) \right].
\end{align}
We then condense the contributions of all harmonics into a single scalar,
\begin{align}
    \mathfrak{T} = \sqrt{\frac{1}{N} \left[ \sum_m G_{\mathrm{cl}}^2(m) + \sum_m G_{\mathrm{q}}^2(m) \right]},
\end{align}
where $N$ is the total number of terms in the two sums. Squaring the $G_{\mathrm{cl/q}}$ prevents contributions of opposite sign from cancelling each other.

To identify the sign of the Chern number, we note that reversing $C$ is equivalent to interchanging the helicities of the drive and the emission, such that the channels are paired as
\begin{align}
    \Delta g^R_R(m, C=1) &\simeq \Delta g^L_L(m, C=-1), \\
    \Delta g^R_L(m, C=1) &\simeq \Delta g^L_R(m, C=-1), \\
    \Delta g^L_R(m, C=1) &\simeq \Delta g^R_L(m, C=-1), \\
    \Delta g^L_L(m, C=1) &\simeq \Delta g^R_R(m, C=-1).
\end{align}

To turn this into a topological response, we define a circular-dichroic $g^{(2)}$ response. First, we sum over the emitted helicity for a given drive helicity and Chern number $C$,
\begin{align}
    \mathcal{S}_C^R(m) & = \Delta g^R_R(m,C) + \Delta g^R_L(m,C), \\
    \mathcal{S}_C^L(m) & = \Delta g^L_R(m,C) + \Delta g^L_L(m,C).
\end{align}
The pairings above imply
\begin{align}
    \mathcal{S}^R_{+1} \simeq \mathcal{S}^L_{-1}, \qquad \mathcal{S}^L_{+1} \simeq \mathcal{S}^R_{-1}.
\end{align}
We therefore define the \textit{circular-dichroic $g^{(2)}$ response}
\begin{align}
    \mathcal{D}_C(m) = \mathcal{S}_C^R(m) - \mathcal{S}_C^L(m),
\end{align}
which changes sign when the sign of the Chern number is reversed,
\begin{align}
    \mathcal{D}_{-C}(m) \simeq -\mathcal{D}_C(m). \label{eq:SM_DChern}
\end{align}
This is confirmed numerically in Table~\ref{tab:D_C_symmetry}, and provides a distinct signature of the transition from $C=+1$ to $C=-1$,
\begin{align}
    C=+1 & \implies \mathcal{D} > 0, \\
    C=-1 & \implies \mathcal{D} < 0.
\end{align}

Finally, a normalized circular-dichroism-type quantity is obtained as
\begin{align}
    \mathfrak{D}_C = \frac{\sum_{m,\,i\in\{R,L\}} \abs{\Delta g^R_i(m,C)}^2 - \sum_{m,\,i\in\{R,L\}} \abs{\Delta g^L_i(m,C)}^2}{\sum_{m,\,i\in\{R,L\}} \abs{\Delta g^R_i(m,C)}^2 + \sum_{m,\,i\in\{R,L\}} \abs{\Delta g^L_i(m,C)}^2},
\end{align}
which sums over the emitted helicity and takes the difference between the $R$ and $L$ drives. By the pairings above, it satisfies
\begin{align}
    \mathfrak{D}_{-C} \simeq -\mathfrak{D}_C,
\end{align}
as required for a dichroism measure.

\begin{table}[htbp]
\centering
\caption{Comparison of $\mathcal{D}_{+1}(m)$ and $\mathcal{D}_{-1}(m)$ with $-\mathcal{D}_{+1}(m)$, confirming $\mathcal{D}_{-C}(m) \simeq -\mathcal{D}_C(m)$ (Eq.~\eqref{eq:SM_DChern}).}
\label{tab:D_C_symmetry}
\begin{tabular}{|c|c|c|c|}
\hline
$m$ & $\mathcal{D}_{+1}(m)$ & $\mathcal{D}_{-1}(m)$ & $-\mathcal{D}_{+1}(m)$ \\
\hline
1 & $-0.0612$ & $0.0612$  & $0.0612$  \\
2 & $-4.7197$ & $4.7197$  & $4.7197$  \\
3 & $9.7932$  & $-9.7932$ & $-9.7932$ \\
4 & $0.8737$  & $-0.8737$ & $-0.8737$ \\
5 & $4.1181$  & $-4.1181$ & $-4.1181$ \\
6 & $0.6644$  & $-0.6644$ & $-0.6644$ \\
7 & $6.8247$  & $-6.8247$ & $-6.8247$ \\
8 & $4.3694$  & $-4.3694$ & $-4.3694$ \\
\hline
\end{tabular}
\end{table}

\end{document}